\documentclass[%
 reprint,
 amsmath,amssymb,
 aps,
 prb,
]{revtex4-2}

\usepackage{graphicx}
\usepackage{dcolumn}
\usepackage{bm}
\usepackage{hyperref}

\begin{document}

\title{Gain-dependent Purcell enhancement, breakdown of Einstein’s relations and superradiance in nanolasers}

\author{Andrey A. Vyshnevyy}

\email{andrey.vyshnevyy@phystech.edu} 
\affiliation{Center for Photonics and 2D Materials, Moscow Institute of Physics and Technology, Dolgoprudny, 141700, Russian Federation}

\date{\today}


\begin{abstract}
Light emitters in a single-mode nanolaser interact with the same cavity field, that gives rise to polarization correlations which transform the cavity mode.
Usually these correlations are ignored, however, collective phenomena can lead to the distinct sub- and superradiance, whose fully quantum description is challenging.
Here, we develop a simple yet rigorous picture of radiative transitions in single-mode nanolasers that accounts for polarization correlations. 
We show that the collective behavior of emitters modifies the photonic density of states leading to gain-dependent Purcell enhancement of spontaneous emission. 
Moreover, the stimulated emission rate is dependent on both the photon number and the laser lineshape. 
As the laser line narrows, stimulated emission becomes stronger than predicted by Einstein's relations and the nanolaser reaches the threshold earlier. 
Finally, we provide concise, ready-to-use expressions for spontaneous and stimulated emission rates seamlessly describing both conventional and superradiant nanolasers.
\end{abstract}

\maketitle
\section{Introduction}
The Purcell effect, first noted in 1946 for radio frequencies~\cite{Purcell1946-pj}, has become a foundation of nanophotonics, playing a pivotal role in surface-enhanced Raman spectroscopy~\cite{Maslovski2019-fk}, ultrafast response dynamics of nanoscale sources of light~\cite{Suhr2010-sy}, and enabling bright and efficient single-photon emitters~\cite{Bennett2005-te,Vyshnevyy2018-npjQI}. The advances in fabrication technology enabled the production of nanolasers, the record small sources of coherent light~\cite{Hill2007-ff,Ding2012-np,Ding2013-if,Nezhad2010-ie,Khajavikhan2012-hv,Prieto2015-jg,wu2015monolayer,gongora2017anapole,Fedyanin2020-nf,tiguntseva2020room,du2020nanolasers,polushkin2020single}, realizing collective coupling of emitters to a single subwavelength-scaled cavity. 
Apart from obvious practical applications, nanolasers attract research interest by their unique and rich physics, including the thresholdless lasing~\cite{Khajavikhan2012-hv,Prieto2015-jg}, delayed and gradual transition to coherent state~\cite{Vyshnevyy2018-sa,Lohof2018-rb}, antibunching~\cite{Chow2014-ze}, and many other phenomena. 
The standard approach to nanolaser modeling employs rate equations~\cite{Yokoyama1989-zt,Bjork1991-ol}, in which the spontaneous and stimulated emission rates originate from Fermi's golden rule~\cite{Casey1978-pp}. 
However, Fermi's golden rule is initially derived for isolated discrete levels not engaged in any interaction except that with the continuum of states. In contrast, electronic states in nanolasers' gain media are disturbed by other charge carriers and phonons, causing their spectral broadening. 
To treat this disturbance rigorously, radiative phenomena must be studied microscopically. 

Using the master equation approach, Ujihara~\cite{Ujihara1993-wh} found the spontaneous emission rate of a single broadened excited emitter into the cavity mode as: 
\begin{equation}\label{sp-FermiRule}
r_\mathrm{sp} = \frac{2\pi |g|^2}{\hbar^2}\int \rho_\mathrm{e}(\omega)\rho(\omega)d\omega, 
\end{equation}
where $\rho_{\rm e}(\omega)$ and $\rho(\omega)$ are Lorentzian densities of states per unit of angular frequency for the emitter and the cavity mode, respectively, and $g$ is the light-matter coupling constant. 
Q. Gu and coauthors obtain a similar expression~\cite{Gu2013-ue}, except that their collisional dephasing model yields a non-Lorentzian density of states for emitters. 
To obtain the stimulated decay rate, one multiplies the spontaneous emission rate by the number of photons in the cavity $N_{\mathrm{p}}$, as dictated by Einstein's relations for radiative transitions:
\begin{equation}\label{st-FermiRule}
r_\mathrm{stim} = N_\mathrm{p}r_\mathrm{sp}. 
\end{equation}

Taking into account that only a fraction of emitters are excited by a factor $n_2(1-n_1)$, where $n_{2(1)}$ are populations of the excited (ground) states, one usually finds the total spontaneous emission rate as: 
\begin{equation}
R_\mathrm{sp} = Nn_2(1-n_1)r_\mathrm{sp}, \label{eq:tot_sp_FGR}   
\end{equation}
where $N$ is the number of emitters in the gain medium. 
Total stimulated emission rate combines the absorption and emission processes in a single term:
\begin{equation}
R_\mathrm{stim} = N(n_2-n_1)r_\mathrm{stim}.\label{eq:tot_stim_FGR}
\end{equation}
In the above expressions $r_\mathrm{sp}$ and $r_\mathrm{stim}$ are defined by Eqs.~(\ref{sp-FermiRule}) and (\ref{st-FermiRule}).

The laser rate equation model incorporating Eqs.~(\ref{eq:tot_sp_FGR}) and (\ref{eq:tot_stim_FGR}) (hereafter abbreviated as REM) shows good agreement with the many-body microscopic model~\cite{Lorke2013-ha}, based on the cluster-expansion approach~\cite{Fricke1996-zi,Gies2007-cy}. 
Although some authors consider the Purcell enhancement independently from the stimulated emission~\cite{Lau2009-wa}, REM currently embodies the established consensus~\cite{Romeira2018-cb}. 
However, Eqs.~(\ref{sp-FermiRule}) and (\ref{st-FermiRule}) treat each quantum emitter independently and ignore their collective behavior.
All emitters interact with the same cavity mode, therefore their small individual polarizations contribute coherently to a macroscopic electric dipole moment.  
This dipole moment affects the dynamics of the laser mode, which, in turn, may significantly modify the radiation rates when the cavity linewidth is comparable or larger than the emitters linewidth~\cite{Leymann2015-nj,Kreinberg2017-pe,Gu2013-ue}. 
This effect is referred to as sub- and superradiance, depending on whether the emission is suppressed or enhanced.

The increasing interest in superradiant lasers is driven by the ability to emit very strong pulses~\cite{jahnke2016giant}, stability of the emission wavelength to fluctuations of the cavity resonance~\cite{bohnet2012steady}, generation of superthermal light~\cite{Kreinberg2017-pe,jahnke2016giant}. Furthermore, most spasers or dielectric nanolasers with highly confined open cavity modes are expected to operate in superradiant regime thanks to low quality factor of the cavity resonance.

Within the semiclassical lasing picture, collective effects are already accounted for by the semiclassical Maxwell-Bloch equations (SMBE), which, as we note below, predict lower threshold population inversion than REM.  
At the same time, SMBE ignore spontaneous emission into the cavity mode because they treat electromagnetic field classically.

A fully quantum description of superradiance in nanolasers is challenging. Existing methods rely on cluster expansion approach~\cite{Leymann2015-nj,Kreinberg2017-pe,jahnke2016giant} or supplement SMBE with the Langevin forces~\cite{andre2019collective,protsenko2021quantum}. 
The former approach can describe any experimental setting, however, it results in a cumbersome set of coupled equations that provide limited insight. 
The latter is mainly applicable to atomic gain media, therefore, ignores the Pauli blocking effects relevant to all nanolasers with semiconductor quantum dots as the source of optical gain. 

Here, we provide a simple picture of radiative transitions in nanolasers, including sub- and superradiance. 
The theory is derived using the rigorous Keldysh formalism for nonequilibrium Green's functions. 
We preserve convenient ``golden rule-like" expressions for the spontaneous and stimulated emission rates, and show that sub- and superradiance in spontaneous emission emerge from the transformation of photonic density of states $\rho(\omega)$ caused by polaritonic transformation (dressing) of the cavity mode. $\rho(\omega)$ depends on the population inversion,  thus, superradiance in nanolasers can be understood as gain-dependent Purcell-enhancement. 
On the other hand, stimulated emission is enhanced by line narrowing, hence the reduced threshold compared to REM and violation of Einstein's relations for single-mode lasers.
Notably, our model shows that sub- and superradiance in nanolasers can appear even without the modification of the collective electronic state, described in the seminal work by R. H. Dicke~\cite{dicke1954coherence}

\section{Nanolaser model}
We consider the Tavis-Cummings model~\cite{Tavis1968-fq} for the interaction of an ensemble of two-level emitters with the single-mode quantized field in a cavity:
\begin{equation}\label{Ham}
H = H_{\mathrm{em}}+H_{\mathrm{cav}}+H_{\mathrm{int}},
\end{equation}
where Hamiltonians of the emitters and the cavity ($\hbar = 1$) are 
$H_{\mathrm{em}} = \sum_{i=1}^N \left( \varepsilon_1 c_{1i}^\dagger c_{1i}+ \varepsilon_2 c_{2i}^\dagger c_{2i} \right)$ 
and $H_{\mathrm{cav}} = \omega_{\rm c} a^\dagger a$, 
correspondingly. 
Here $\varepsilon_1,\,\varepsilon_2$ are the energies of the ground and excited emitter states, $\omega_{\rm c}$ is the central frequency of the cavity mode, $c_{1i}^{(\dagger)}, c_{2i}^{(\dagger)}, a^{(\dagger)}$ are annihilation (creation) operators of the ground and excited levels of the \textit{ith} emitter and the cavity mode, respectively, $N$ is the number of emitters in the cavity.
Light-matter interaction is governed by $H_{\mathrm{int}} = \sum_{i=1}^N (g_i c_2^\dagger c_1 a + h.c.)$, where $g_i$ is the constant of light-matter coupling. 
To achieve lasing, one should pump the excited electronic state and collect electrons from the ground state, which in our model is realized by connecting each emitter's electronic levels to corresponding fermionic reservoirs, as shown in Fig.~1a. 
The cavity loss is modeled as coupling to a bosonic bath. 
Here, we do not impose restriction $c_1^\dagger c_1 + c_2^\dagger c_2 = 1,$ and allow electrons to populate the excited and ground levels independently, which applies to gain media based on semiconductor quantum dots~\cite{Gies2007-cy}. 
Also, the absence of the restriction allows for a more straightforward diagrammatic treatment than otherwise~\cite{Shchadilova2020-mq}. 

\begin{figure}[htbp]
\centering
\includegraphics[width=7cm]{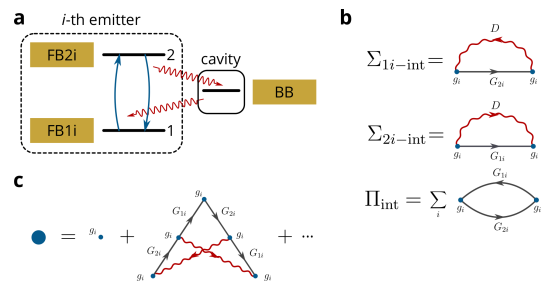}
\caption{(a) Schematic of the laser model comprising N two-leveled emitters radiatively coupled to the cavity mode. 
Ground (labeled 1) and excited (labeled 2) levels of each emitter are coupled to fermionic reservoirs (labeled FB1i and FB2i for the \textit{ith} emitter) responsible for incoherent pumping and collection of electrons and homogeneous broadening of electronic levels. 
Cavity loss is modeled by coupling to bosonic bath BB. (b) Self-energies originating from light-matter interaction in the system for the cavity photons $\Pi_{\mathrm{int}}$, ground $\Sigma_{1i-{\mathrm{int}}}$ and excited $\Sigma_{2i-{\mathrm{int}}}$ electronic levels. (c) Vertex function expansion showing the leading-order correction.}\label{fig:1}
\end{figure}

We study the described driven-dissipative Tavis-Cummings model using the Keldysh diagrammatic technique~\cite{Keldysh:1964ud} for non-equilibrium Green's functions~\cite{stefanucci_van_leeuwen_2013}. Within this method all observables are defined through the Green's functions of electronic and photonic states, while the emission rates follow directly from quantum kinetic equations. 
Importantly, we can account for large dipole moment of the gain medium in all perturbation orders via the partial summation of relevant diagrams.

Couplings to the fermionic reservoirs are described by the retarded and lesser self-energies~\cite{Arseev2015-he} $\Sigma_{\mathrm{FB1,2i}}^R(\omega)=-i\gamma_{1,2}/2$ and $\Sigma^<_{\mathrm{FB1,2i}}(\omega) = in_{1,2}^0\gamma_{1,2}$, whereas similar self-energies for the coupling to a bosonic bath are $\Pi_{\mathrm{BB}}^R(\omega)=-i\kappa/2$ and, neglecting thermal occupancy of the bosonic bath, $\Pi_{\mathrm{BB}}^<(\omega) = 0$. 
Here $\kappa, \gamma_1, \gamma_2$ are the photonic mode’s damping rate, and tunneling rates to the ground and excited levels, correspondingly. 
Superscripts $<,>,R,A$ denote the lesser, greater, retarded and advanced components of Green's functions and self-energies. 
For simplicity, we let populations $n_{1,2}^0$ of FB1i and FB2i be energy-independent. 
We focus on the steady-state solution, thus for any quantity $O$: $O^A(\omega)=\left[O^R(\omega)\right]^*$.
Also, $O^> = O^<+O^R-O^A$. 

Figure~1b depicts self-energies from light-matter interaction.
We treat the interaction self-consistently, which means that all lines in diagrams in Fig.~1b label dressed Green’s functions. 
Our approximation is valid when the number of photons in the cavity is not too large so that the corrections to the vertex function, shown in Fig.~1c, can be neglected. 
The components of interaction-induced polarization operator read:
\begin{eqnarray}\label{eq:Pi^R}
\Pi_\mathrm{int}^R(\omega) =&& -i\sum_{i=1}^N |g_i|^2\int\frac{d\omega'}{2\pi}\left[G_{1i}^<(\omega')G_{2i}^R(\omega'+\omega)\right.\\ \nonumber
&&\left.+G_{1i}^A(\omega')G_{2i}^<(\omega+\omega')\right],\\ 
\Pi_\mathrm{int}^\lessgtr(\omega) =&& -i\sum_{i=1}^N |g_i|^2\int\frac{ d\omega'}{2\pi}G_{1i}^\gtrless(\omega')G_{2i}^\lessgtr(\omega'+\omega),
\end{eqnarray}
where $G_{1,2i}$ denote Green’s functions of the \textit{ith} emitter’s electronic levels. 
To derive the emission rates, we use the Kadanoff-Baym (quantum rate) equations. In the stationary state:
\begin{eqnarray}
\nonumber
0 =&& \frac{\partial N_\mathrm{p}}{\partial t}= \int \frac{d\omega}{2\pi}\left[\Pi^R(\omega)-\Pi^A(\omega)\right]D^<(\omega)\\ 
\label{eq:Kadanoff-Baym}&&+ \int \frac{d\omega}{2\pi}\Pi^<(\omega)\left[D^A(\omega)-D^R(\omega)\right],
\end{eqnarray}
in which $N_{\rm p}$ is the number of photons and $D(\omega)$ is the Fourier transform of the photonic Green’s function $D(t,t')=-i\left<\mathcal{T}_K\left[ a(t) a^\dagger(t')\right]\right>$, where $\mathcal{T}_K$ denotes time ordering along the Keldysh contour. 
The polarization operator contains both the dissipative and interaction parts $\Pi=\Pi_\mathrm{BB}+\Pi_\mathrm{int}$. 
In this rate equation, the total stimulated and spontaneous emission rates into the cavity mode are given by:
\begin{eqnarray}
R_\mathrm{stim}^\mathrm{tot} =&& \int \frac{d\omega}{2\pi} \left[\Pi^R_\mathrm{int}(\omega)-\Pi^A_\mathrm{int}(\omega)\right]D^<(\omega), \\
R_\mathrm{sp}^\mathrm{tot} =&& \int \frac{d\omega}{2\pi} \Pi^<_\mathrm{int}(\omega)\left[D^A(\omega)-D^R(\omega)\right].
\end{eqnarray}

The physical meaning of these equations becomes clear after we introduce photonic $\rho(\omega)=(-1/\pi)\mathrm{Im}\left[D^R(\omega)\right]$ and electronic $\rho_{1,2i}(\omega)=(-1/\pi)\mathrm{Im}\left[G_{1,2i}^R(\omega)\right]$ spectral functions (densities of states) and corresponding non-equilibrium populations $n(\omega)$ and $n_{1,2i}(\omega)$:
$ D^<(\omega) = -2\pi in(\omega)\rho(\omega),\, G_{1,2i}^<(\omega) = 2\pi in_{1,2i}(\omega)\rho_{1,2i}(\omega). 
$
Using the identity $\Pi^R-\Pi^A = \Pi^>-\Pi^<$, we express the spontaneous and stimulated emission rates of the \textit{ith} emitter as:
\begin{eqnarray}
\nonumber r_\mathrm{i-sp} =&& 2\pi|g_i|^2\int d\omega d\omega' n_{2i}(\omega+\omega')[1-n_{1i}(\omega')]\\
&&\label{eq:i-sp-general}\times\rho_2(\omega+\omega')\rho_1(\omega')\rho(\omega),\\
\nonumber r_\mathrm{i-stim} =&& 2\pi|g_i|^2\int d\omega d\omega' [n_{2i}(\omega+\omega')-n_{1i}(\omega')]\\
&&\label{eq:i-stim-general}\times \rho_2(\omega+\omega')\rho_1(\omega')n(\omega)\rho(\omega) 
\end{eqnarray}

The obtained equations are very similar to those for the rates of interband transitions in bulk semiconductors~\cite{Casey1978-pp}. 
Note that Eqs.~(\ref{eq:i-sp-general}) and (\ref{eq:i-stim-general}) treat initially discrete bound electronic levels and the cavity mode as if they are energy bands with energy-dependent populations and densities of states. 
Transformation of the discrete level into an energy band can be qualitatively explained by its interaction with the continuum of reservoir modes, producing a continuum of hybridized ``modes of the Universe" with different coupling strengths to the quantum emitter.
Energy-dependent populations and densities of states absorb non-Markovian many-body effects into convenient ``golden-rule-like" form. 
Pumping and dephasing, as well as radiative transitions themselves, self-consistently determine the densities of states and population functions. 
The non-equilibrium photonic population is dictated by the balance of the \textit{spectral densities} of emission and loss rates in Eq.~(\ref{eq:Kadanoff-Baym}), leading to:
\begin{equation}\label{eq:n(omega)}
n(\omega) = \Pi^</(\Pi^R-\Pi^A).
\end{equation}
Finally, the retarded Green’s function required for the calculation of the photonic density of states is given by the Schwinger-Dyson equation:
\begin{equation}\label{eq:D^R}
\left[D^R(\omega)\right]^{-1} = \omega-\omega_{\rm c}-\Pi^R(\omega).
\end{equation}
Similar equations determine the densities of states of electronic levels $\rho_1(\omega)$, $\rho_2(\omega)$ and their frequency-dependent populatons $n_1(\omega)$, $n_2(\omega)$. 
In addition to the transformation of the cavity mode, the full set of self-consistent equations describes non-linear optical response of the gain medium which includes population-related phenomena, such as gain saturation and spectral hole burning, and mixing of electronic levels via the strong cavity field usually named AC Stark effect. 
For the sake of simplicity, we first focus on the transformation of the cavity mode while neglecting nonlinearities in the gain medium. 
\section{Nanolaser in the limit of a linear gain medium}

We assume that all coupling constants are equal $g_i=g$ and ignore the transformations of the emitters levels and densities of states by the cavity field.
Mathematically this corresponds to the limit of $g\rightarrow 0$, $N\rightarrow \infty$ while $|g|^2N$ is constant. 
Such a limit matches SMBE~\cite{Narducci1988-hb}. 
For any finite number of photons in the cavity, we can ignore the contributions from light-matter interaction to total self-energies of the electronic states $\Sigma_{1,2i} \approx \Sigma_\mathrm{FB1,2i}$ and, more importantly, vertex corrections (Fig.~1c). 
As a result, the electronic densities of states are the usual Lorentzian functions 
\begin{equation}
\rho_{1,2}(\omega) = \frac{\gamma_\mathrm{1,2}/(2\pi)}{(\omega-\varepsilon_{1,2})^2+\left(\gamma_\mathrm{1,2}/2\right)^2}
\end{equation}
while the populations of electronic levels coincide with those of the fermionic reservoirs $n_{1,2}(\omega) = n_{1,2}^0$. 
However, we should account for interaction in polarization operator $\Pi = \Pi_\mathrm{BB} + \Pi_\mathrm{int}$. 
Using Eq.~(\ref{eq:Pi^R}), we find the retarded interaction-induced polarization operator responsible for the polaritonic transformation of the cavity mode:
\begin{equation}
\Pi_\mathrm{int}^R(\omega) = -\frac{|g|^2N(n_2^0-n_1^0)}{\omega-\omega_{21}+i\gamma_\mathrm{e}/2},
\end{equation}
where $\omega_{21} = \varepsilon_2 - \varepsilon_1$ and $\gamma_\mathrm{e}=\gamma_1+\gamma_2$.
Light-matter interaction produces a term that is proportional to the classical electric susceptibility of the gain medium. 
The retarded Green’s function, evaluated from Eq.~(\ref{eq:D^R}), gives the photonic density of states modified by the gain medium:
\begin{eqnarray}
\rho(\omega) = -\frac{1}{\pi}\mathrm{Im}\left[\omega-\omega_{\rm c}+i\frac{\kappa}{2}+\frac{|g|^2N(n_2^0-n_1^0)}{\omega-\omega_{21}+i\gamma_\mathrm{e}/2}\right]^{-1}\label{eq:rho}
\end{eqnarray}
It is worth noting that, within the studied limit, the photonic Green's function is the Green's function of SMBE:
\begin{eqnarray}
\nonumber id_t D^R=&& (\omega_{\rm c}-i\kappa/2)D^R+g^*P+\delta(t),\\
\label{eq:SMBE}id_t P =&& (\omega_{21}-i\gamma_\mathrm{e}/2)P+gD^RN(n_2^0-n_1^0)
\end{eqnarray}
where $P$ denotes the dipole moment of gain medium.

Next, instead of the photonic population, we directly extract the lineshape function $S(\omega)=n(\omega)\rho(\omega)$ by balancing photon emission and loss, which gives:
\begin{equation}\label{eq:photon_rate}
\kappa S(\omega) = R_\mathrm{sp}(\omega) + \mathcal{G}(\omega)S(\omega),
\end{equation}
where $R_\mathrm{sp}(\omega) = 2\pi N|g|^2 n_2^0(1-n_1^0) \rho_{\rm e}(\omega)\rho(\omega)$ is the rate of spontaneous emission of all emitters per unit of angular frequency, and $\mathcal{G}(\omega) = 2\pi N |g|^2 (n_2^0-n_1^0) \rho_\mathrm{e}(\omega)$ is the modal gain spectrum, with $\rho_\mathrm{e}(\omega)=\int d\omega' \rho_1(\omega')\rho_2(\omega+\omega')$ being the emitter's density of states. 
Once $S(\omega)$ is determined, we can calculate the spontaneous and stimulated emission rates as:
\begin{equation}\label{eq:sp-stim_weak}
R_\mathrm{sp}^\mathrm{tot} = \int R_\mathrm{sp}(\omega)d\omega,\qquad	R_\mathrm{stim}^\mathrm{tot} = \int \mathcal{G}(\omega)S(\omega)d\omega.
\end{equation}
These equations are very similar to Eq.~(\ref{eq:tot_sp_FGR}) and (\ref{eq:tot_stim_FGR}). 
In fact, one can recover Eq.~(\ref{eq:tot_sp_FGR}) and (\ref{eq:tot_stim_FGR}) from Eq.~(\ref{eq:sp-stim_weak}) by manually setting up
\begin{equation}
    \rho(\omega)=\frac{\kappa/(2\pi)}{(\omega-\omega_\mathrm{c})^2+(\kappa/2)^2}
\end{equation} 
and $S(\omega)=N_{\rm p}\rho(\omega)$. 
This means that observable characteristics of lasing, such as linewidth narrowing and frequency pulling, are the result of the cavity mode transformation under the collective dipole moment of emitters. 
Importantly, this transformation occurs in any structure capable of lasing, regardless of the number of emitters and the coupling strength between the emitters and the cavity mode. 
Thus, fundamentally, lasers always operate beyond the weak coupling regime.

\begin{figure}[htbp]
\centering
\includegraphics[width=7cm]{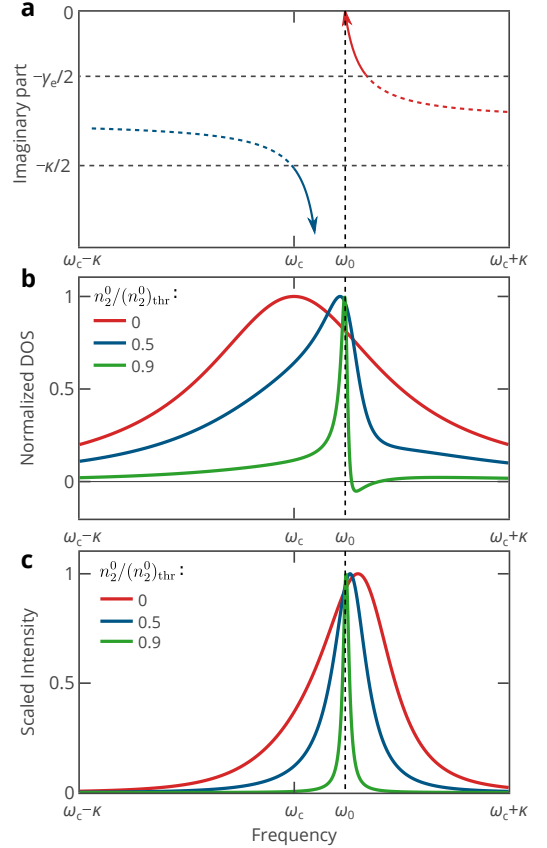}%
\caption{(a) Poles of the retarded photonic Green's function as a function of population inversion. Arrows show the direction of increase in inversion. Dashed parts of the curves are relevant to negative inversion $n_2^0 < n_1^0$, i.e., absorbing gain medium.  (b) The optical density of states $\rho(\omega)$, normalized to its maximum value and (c) the normalized emission spectrum plotted at different degrees of population inversion. All calculations assume $\Delta = \kappa/3,\, \gamma_\mathrm{e}=0.4 \kappa,\, n_1^0=0$. $(n_2^0)_\mathrm{thr}$ denotes the threshold population of the excited level, given by Eq.~(\ref{eq:threshold})}\label{fig:2}
\end{figure}

To determine the laser threshold, we analyze the poles of $D^R(\omega)$ shown in Fig.~2a. They obey the equation 
\begin{equation}
   \left(\omega-\omega_{\rm c}+i\kappa/2\right)\left(\omega-\omega_{21}+i\gamma_\mathrm{e}/2\right)+|g|^2 N(n_2^0-n_1^0)=0,
\end{equation} solving which, we find that one of the poles crosses the real axis at $\omega_0 = \omega_{\rm c} + \kappa \Delta/(\kappa+\gamma_\mathrm{e}) $ when $\mathcal{G}(\omega_0)=\kappa$, or, in terms of the population inversion, when 
\begin{equation}
    n_2^0-n_1^0 = \frac{\gamma_\mathrm{e}\kappa}{4N|g|^2}\left[1+\left(\frac{2\Delta}{\kappa+\gamma_\mathrm{e}}\right)^2\right].\label{eq:threshold}
\end{equation} 
Above, $\Delta=\omega_{21}-\omega_\mathrm{c}$ is the detuning between the emitters and the cavity.
Notably, our nanolaser model predicts a different lasing threshold from REM. The latter reaches the balance between stimulated emission and the photon loss 
$2\pi N|g|^2(n_2^0-n_1^0)N_\mathrm{p}\int \rho_\mathrm{e}(\omega)\rho(\omega)d\omega = \kappa N_\mathrm{p}$ at a population inversion of 
\begin{equation}
    n_2^0-n_1^0 = \frac{(\gamma_\mathrm{e}+\kappa)\kappa}{4N|g|^2}\left[1+\left(\frac{2\Delta}{\kappa+\gamma_\mathrm{e}}\right)^2\right],\label{eq:threshold_REM}
\end{equation} 
exceeding the value in Eq.~(\ref{eq:threshold}) by factor of $(\gamma_{\rm e}+\kappa)/\gamma_{\rm e}$. At the same time, Eq.~(\ref{eq:threshold}) agrees with SMBE, while the threshold of REM is not~\cite{Narducci1988-hb}. Unlike REM, SMBE include the collective dipole moment of emitters classically, thus the obtained agreement can be understood as a particular case of quantum-classical correspondence.

Figure~2b,c shows how the optical density of states and the lineshape function evolve with the increase in population inversion. For illustrative purposes, we assumed $\Delta = \kappa/3,\, \gamma_\mathrm{e}=0.4 \kappa,\, n_1^0=0$. Remarkably, the optical density of states is negative in a limited spectral band corresponding to $\mathcal{G}(\omega)>\kappa$. Although negative densities of states look exotic and are impossible in purely lossy environments, they have been reported for optical systems which combine amplifying and absorptive media~\cite{Hughes_PRL_2021}. At the same time, no exotic behavior is seen in the calculated lineshape function $S(\omega)$ which remains positively-valued and exhibits peak narrowing and frequency pulling. As the population inversion approaches its threshold value, the laser luminescence peak approaches $\omega_0$, which again fully agrees with SMBE~\cite{Narducci1988-hb}.


Although the photonic density of states and the lineshape function have complicated profiles, it is possible to calculate the total spontaneous and stimulated emission rates analytically. 
After some algebra exploiting the residue theorem (see the Supplement 1, Section I and II for details), we obtain one of key results of this paper:
\begin{eqnarray}
R_\mathrm{sp}^\mathrm{tot} =&& n_2^0(1-n_1^0)\frac{N|g|^2}{\hbar^2}\frac{\kappa+\gamma_\mathrm{e}-\mathcal{G}_\mathrm{max}/2}{\Delta^2+\left(\frac{\kappa+\gamma_\mathrm{e}-\mathcal{G}_\mathrm{max}/2}{2}\right)^2},\label{eq:sp-an}\\
R_\mathrm{stim}^\mathrm{tot}=&&\kappa N_\mathrm{p}-R_\mathrm{sp}^\mathrm{tot}\label{eq:stim-an},
\end{eqnarray}
where the number of photons in the cavity is
\begin{equation}\label{eq:N_p-an}
N_\mathrm{p} = \int S(\omega)d\omega = \frac{\frac{4N|g|^2}{\hbar^2} \frac{n_2^0(1-n_1^0)}{\kappa+\gamma_\mathrm{e}}}{\kappa\left[1+\left(\frac{2\Delta}{\kappa+\gamma_\mathrm{e}}\right)^2\right]-\mathcal{G}_\mathrm{max}}, 
\end{equation}
and $\mathcal{G}_\mathrm{max}=\mathcal{G}(\omega_{21})=4N(|g|^2/\hbar^2)(n_2^0-n_1^0)/\gamma_\mathrm{e}$. In the above expressions we have restored $\hbar$ for the convenience.

The effect of the polaritonic transformation on the total spontaneous emission rate is equivalent to the reduction of the “effective” cavity damping rate.
As a result, Eq.~(\ref{eq:sp-an}) predict the increase in the total spontaneous emission rate, i.e.,  superradiance, at positive population inversion ($G_\mathrm{max} > 0$) and subradiance when the gain medium absorbs light ($G_\mathrm{max} < 0$), unless $\Delta$ is very large. 
This leads to gain-dependent values of beta and Purcell factors even in the absence of inhomogeneous broadening. In the case when $\Delta = 0$, the Purcell factor increases by a factor of 2 as the population inversion approaches the threshold value. 
Rates of the spontaneous and stimulated emission of radiation are connected by the Einstein relations, which in the case of a single cavity mode is typically expressed by Eq.~(\ref{st-FermiRule}). 
However, in our case, Eq.~(\ref{st-FermiRule}) holds only for \textit{spectral densities} of the spontaneous and stimulated emission rates of an excited emitter $r_\mathrm{stim}(\omega) = r_\mathrm{sp}(\omega)n(\omega)$, while the total rates violate it due to nonuniform $n(\omega)$, as seen from Eqs.~(\ref{eq:sp-an}) and ~(\ref{eq:stim-an}). 

To demonstrate violation of the Einstein's relations, we plot the violation factor defined as 
\begin{equation}
F_V = \frac{R_\mathrm{stim}}{N_{\rm p} R_\mathrm{sp}}\frac{n_2^0 -n_1^0}{n_2^0(1-n_1^0)}\label{eq:violation}
\end{equation}
as a function of the photon number (Fig.~3). 
According to this definition, $F_V = 1$ for emission rates given by Eqs.~(\ref{eq:tot_sp_FGR}) and~(\ref{eq:tot_stim_FGR}) which obey Einstein's relation. 
By contrast, the violation factor calculated using Eqs.~(\ref{eq:sp-an}) and~(\ref{eq:stim-an}) reaches above 2.75. 
Note, that, even when $N_{\rm p} \sim 0$, i.e., when the photonic density of states is still Lorentzian, the $F_V$ is still above 2.5, thanks to nonuniform $n(\omega)$. 

\begin{figure}[htbp]
\centering
\includegraphics[width=7cm]{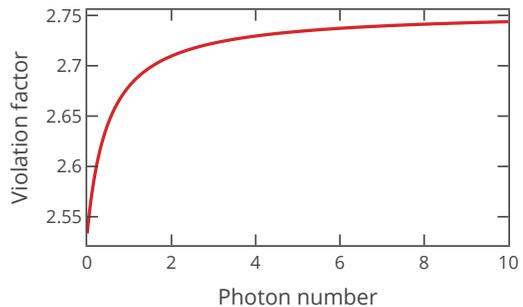}%
\caption{Einstein's relation violation factor determined by Eq.~(\ref{eq:violation}) as a function of the photon number in the laser mode. 
The nanolaser parameters are the same as in Fig.~2.}\label{fig:3}
\end{figure}

Although our equations predict the violation of Einstein's relations, this violation becomes weak if $\kappa \ll \gamma_\mathrm{e}$, or the nanolaser belongs to ``class-B" lasers~\cite{arecchi1984deterministic}. This is not the result of weak coupling, but rather is caused by insensitivity of emission rates to the transformations of $\rho(\omega)$ and $S(\omega)$. 
In fact, one can simply assume $\rho(\omega) = \delta(\omega-\omega_{\rm c})$ and $S(\omega) = N_{\rm p} \delta(\omega-\omega_{\rm c})$. 
At the same time, solid-state nanolasers can disobey $\kappa \ll \gamma_\mathrm{e}$ since the quality factor of wavelength-scaled nanocavities can be as small as few hundred~\cite{Ding2012-np} or even less for open cavities while $\gamma_\mathrm{e}$ of quantum dots can be made as small as 10 $\mathrm{\mu eV}$ at low temperature~\cite{kuhlmann2015transform}. In such case, they are often referred to as superradiant, with a number having been experimentally realized~\cite{lermer2013high,assmann2010ultrafast}. 

\section{Feedback, lasing threshold and limitations of the theory}

In this section, we briefly discuss the role of the cavity feedback on electronic levels and limitations of the described theory. 
As the population inversion increases, the number of photons in the cavity grows, eventually making light-matter-interaction-related contributions $\Sigma_{1i-\mathrm{int}}, \Sigma_{2i-\mathrm{int}}$ to the electronic self-energies large. 
To estimate the photon number, when we can no longer neglect these contributions, we directly calculate $\Sigma_{1i-\mathrm{int}}^R$. Assuming $N_\mathrm{p} \gg 1$, the lineshape can be approximated as $S(\omega) = N_\mathrm{p}\delta(\omega-\omega_0)$, or, equivalently, $D^<(\omega)=-2\pi i N_{\rm p} \delta(\omega-\omega_0)$. 
Neglecting a small contribution from $D^R(\omega)$ we find
\begin{eqnarray}
 \nonumber\Sigma_{1i-\mathrm{int}}^R \approx&& \; i|g|^2\int D^<(\omega')G_{2i}^R(\omega+\omega')\frac{d\omega'}{2\pi}\approx \\ &&\frac{|g|^2N_\mathrm{p}}{\omega+\omega_0-\varepsilon_2+i\gamma_2/2}.      
\end{eqnarray}
Demanding $|\Sigma_{1i-\mathrm{int}}^R(\omega)|\ll |\Sigma_{FB1i}^R(\omega)|$ for real frequencies, we obtain $N_\mathrm{p} \ll \gamma_1\gamma_2/(4|g|^2)$. When the number of photons approaches this limit, two phenomena occur. 
Firstly, due to the rapid stimulated emission, populations of electronic levels $n_1(\omega), n_2(\omega)$ start to deviate from fermionic bath’s occupancies $n_1^0, n_2^0$, which results in gain saturation and spectral hole burning. 
Secondly, the corrections to the electronic densities of states make the modal gain spectrum $\mathcal{G}(\omega)=2\pi N|g|^2 (n_2-n_1)\rho_\mathrm{e}(\omega)$ explicitly dependent on the number of photons in the cavity mode, even if $n_1$ and $n_2$ are maintained at a constant level.
This regime corresponds to the onset of the AC Stark effect~\cite{AC-StarkQD}. 
In most nanolasers, only a small fraction of emitters’ homogeneous broadening is caused by the pumping or decay, with the rest of it being due to elastic phase breaking processes, thus gain saturation occurs much earlier than the AC Stark effect. 
In this case, $n_{1(2)}^0$ in Eqs.~(\ref{eq:sp-an}) and (\ref{eq:N_p-an}) are to be replaced with $n_{1(2)}=\langle c_{1(2)}^\dagger c_{1(2)}\rangle$, which should be obtained from  $n_{1(2)} = n_{1(2)}^0 \pm \kappa N_{\rm p}/(N\gamma_{r1(r2)})$, where ``+" corresponds to the equation for $n_1$, while $\gamma_{r1(r2)}$ are population relaxation rates, now different from electronic level homogeneous broadenings $\gamma_{1(2)}$.  

Next, we demonstrate that our theory describes nanolasers in both the lasing and nonlasing regimes by estimating the lasing threshold and showing that it is within limits of applicability of the developed theory.
The concept of lasing  and, in particular, lasing threshold is attracting much attention in recent years. 
With the experimental demonstration of so-called thresholdless nanolasers~\cite{Khajavikhan2012-hv}, the usual definition based on features of input-output (light-light or light-current) characteristics becomes ill-defined~\cite{ning2013laser}. 
As a result, alternative definitions have been proposed~\cite{Chow2014-ze, Vyshnevyy2018-sa, vyshnevyy2020elusive, Lohof2018-rb}, most of which are based on the statistics of radiation, which is Poissonian for coherent light, while that of incoherent, or thermal, light is described by the Bose-Einstein distribution.
We consider a two-particle photonic Green's function $D^\mathrm{2p}(t_1, t_2, t_3, t_4)=-\langle \mathcal{T}_K[a(t_1)a(t_2)a^\dagger(t_3)a^\dagger(t_4)] \rangle$. 
This function is directly related to the second-order coherence function $g^{(2)}(\tau) = \frac{\langle a^\dagger(t)a^\dagger(t+\tau)a(t+\tau)a(t) \rangle}{\langle a^\dagger(t)a(t)\rangle\langle a^\dagger(t+\tau)a(t+\tau)\rangle}$, which is capable of unambiguous distinction between thermal ($g^{(2)}(0)=2$) and coherent ($g^{(2)}(0)=1$) radiation. 
Fig.~4 shows the diagrammatic series for $D^\mathrm{2p}$. 
The first two terms correspond to:
\begin{eqnarray}
    \nonumber\langle a^\dagger(t)a^\dagger(t+\tau)a(t+\tau)a(t) \rangle \approx \\
    \nonumber\langle a^\dagger(t)a(t)\rangle\langle a^\dagger(t+\tau)a(t+\tau)\rangle + \\
    \langle a^\dagger(t)a(t+\tau)\rangle\langle a^\dagger(t+\tau)a(t)\rangle,
\end{eqnarray}
which translates into the Siegert relation~\cite{siegert1943fluctuations} $g^{(2)}(\tau)=1+|g^{(1)}(\tau)|^2$, where $g^{(1)}(\tau)$ is the first-order coherence function.
Radiation obeying the Siegert relation is obviously thermal ($g^{(2)}(0)=2$) since $g^{(1)}(\tau=0)=1$ for a single-mode electromagnetic radiation. 
The breakdown of the Siegert relation happens when the third and forth terms of the diagrammatic series become comparable to the first two, that is
\begin{equation}
    \frac{|g|^4 N N_{\rm p}^4}{\gamma_1 \gamma_2 \gamma_\mathrm{e} \gamma_\mathrm{line}} \sim N_{\rm p}^2,\label{eq:Np_coh_thr}
\end{equation}
where $\gamma_\mathrm{line} \ll \gamma_\mathrm{e}$ is the FWHM of the laser peak (see Supplement 1, section III for more details). 

\begin{figure}[htbp]
\centering
\includegraphics[width=7cm]{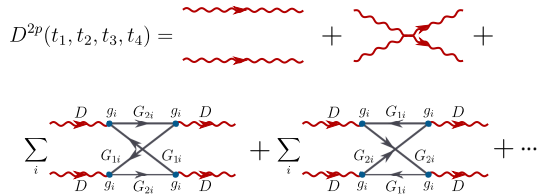}
\caption{Diagrammatic series for the two-particle Green's function for photons.}\label{fig:4}
\end{figure}

To determine the limits of applicability of the employed self-consistent second order approximation we roughly estimate the previously discarded vertex correction (Fig.~1c). 
Its relative value in the leading order does not exceed $|\delta g/g| \sim \left[ 4|g|^2N_\mathrm{p}/(\gamma_1\gamma_2)\right]^2$, setting up the limitation on the photon number $N_\mathrm{p}\ll \gamma_1\gamma_2/(4|g|^2)$ (see Supplement 1, section IV). 
When the number of photons is moderately high, the vertex correction can still be neglected compared to the corrections to the electronic Green’s functions since the former is proportional to the second power of small parameter $4|g|^2N_\mathrm{p}/(\gamma_1\gamma_2)$ while the latter to the first power of this parameter. 
Notably, the photon number at the coherence threshold, determined by Eq.~(\ref{eq:Np_coh_thr}) is well within this limit, thus the developed theory applies both to non-lasing and lasing regimes. 
Eventually, when $4|g|^2N_\mathrm{p}/(\gamma_1\gamma_2) \sim 1$, the nanolaser enters mostly unexplored regime as SMBE, REM and this theory no longer describe the physics of the nanolaser, even if $\kappa \ll \gamma_\mathrm{e}$.

\section{Conclusion}

To summarize, we present a systematic study of lasing nanocavities with continuously pumped gain medium using the Keldysh formalism for non-equilibrium Green’s functions. 
Our theory incorporates non-Markovian cavity-emitter dynamics and the polarization correlations between emitters into population functions and densities of states (see Eqs.~(\ref{eq:i-sp-general}) and (\ref{eq:i-stim-general})). 
We show that nanolasers always operate beyond the weak coupling limit, as the photonic mode is transformed by the collective electric dipole moment of the emitters in the active region, which leads to sub- and superradiance.
Remarkably, the discovered mechanism of sub- and superradiance is fundamentally different from the Dicke superradiance since no transformation of the electronic state is involved.
Key equations of our theory~(\ref{eq:photon_rate},\ref{eq:sp-stim_weak}) have a simple rate equation structure. 
Our approach systematically adds spontaneous emission to semiclassical Maxwell-Bloch equations.
The use of the Keldysh formalism allowed us to maintain a sufficient level of mathematical rigor by revealing the limits of validity for all involved approximations and the error scales associated with them.

Remarkably, while the derivation relies on the Keldysh formalism, the application of the developed theory to nanolasers at $N_\mathrm{p}\ll \gamma_1\gamma_2/(4|g|^2)$ does not require mastery of many-body perturbation theory. The retarded Green's function of photons, which describes the transformation of the photonic density of states, can be obtained directly from Maxwell-Bloch equations~(\ref{eq:SMBE}) or classical electrodynamics. The laser lineshape naturally emerges from the rate equation~(\ref{eq:photon_rate}). Computations of input-output curves would benefit from compact ready-to-use expressions~(\ref{eq:sp-an},\ref{eq:stim-an},\ref{eq:N_p-an}) connecting the population inversion to the number of photons and radiative emission rates. These expressions shed light on nanolaser physics by revealing gain-dependent Purcell enhancement and breakdown of Einstein’s relations due to line narrowing. Also, we predict the negative spectral density of spontaneous emission near the lasing threshold. 
Finally, we discuss the nonlinear optical response of the gain medium and show that the transition to coherent emission happens within the validity limits of our theory.

With foreseeable extensions to broadband gain media, such as bulk semiconductors and quantum wells, our theory creates a firm ground for studies of many-body phenomena in active nanophotonic devices.

\begin{acknowledgments}
The author was supported by the stipend from the President of the Russian Federation (SP-1125.2021.5), the Ministry of Science and Higher Education of the Russian Federation (Agreement No. 075-15-2021-606). The author is grateful to Alex Krasnok, Sergey Lepeshov, Denis Baranov and Igor Khramtsov for stimulating discussions.
\end{acknowledgments}

\end{document}